\documentclass[reqno]{amsart}
\usepackage{bm}
\newcommand{\Rset}{\mathbb{R}}
\newcommand{\Cset}{\mathbb{C}}
\newcommand{\half}{{\textstyle\frac{1}{2}}}

\newtheorem{theorem}{Theorem}[section]
\newtheorem{proposition}[theorem]{Proposition}
\newtheorem{definition}[theorem]{Definition}

\theoremstyle{remark}
\newtheorem{example}[theorem]{Example}

\newcommand{\hreals}{\hat{\Rset}}
\newcommand{\hcnums}{\hat{\Cset}}
\newcommand{\cnums}{\Cset}

\newcommand{\Vi}{V_{\mathrm{i}}}

\newcommand{\lspan}{\operatorname{span}}

\newcommand{\cG}{\mathcal{G}}

\newcommand{\hp}{\hat{p}}

\newcommand{\hz}{\hat{z}}

\newcommand{\hu}{\hat{u}}

\newcommand{\hell}{\hat{\ell}}
\newcommand{\hatm}{\hat{m}}

\newcommand{\rA}{\mathrm{A}}
\newcommand{\rB}{\mathrm{B}}
\newcommand{\rC}{\mathrm{C}}

\newcommand{\sA}{{\scriptscriptstyle \rA}}
\newcommand{\sB}{{\scriptscriptstyle \rB}}

\newcommand{\subp}[1]{_{#1}}
\newcommand{\supp}[1]{^{#1}}

\newcommand{\du}[2]{_{#1}^{\;#2}}
\newcommand{\ud}[2]{^{#1}{}_{#2}}

\newcommand{\spPK}{\mathbb{PK}}

\newcommand{\spRPK}{\Rset\mathbb{PK}}
\newcommand{\spS}{\mathbb{S}}

\newcommand{\lp}{\left(}
\newcommand{\rp}{\right)}

\newcommand{\tT}{\mathbf{T}}

\newcommand{\vk}{\bm{\mathit{k}}}
\newcommand{\vell}{{\bm{\ell}}}
\newcommand{\vm}{{\bm{\mathit{m}}}}
\newcommand{\vn}{{\bm{\mathit{n}}}}
\newcommand{\hvell}{\hat{\vell}}
\newcommand{\hvm}{\hat{\vm}}
\newcommand{\hvn}{\hat{\vn}}

\newcommand{\cW}{\mathcal{W}}

\newcommand{\bwt}{\mathrm{b}}
\newcommand{\bo}{\bwt}
\newcommand{\maxbo}{\bwt_{\max}}

\begin{document}

\title{Alignment and the classification of Lorentz-signature tensors.}
\author{R. Milson}
\address{Dept. Mathematics and Statistics\\
  Dalhousie University\\
  Halifax, Nova Scotia B3H 3J5 \\
  Canada \\
  E-mail: rmilson@dal.ca}

\begin{abstract}
  We define the notion of an aligned null direction, a
  Lorentz-signature analogue of the eigenvector concept that is valid
  for arbitrary tensor types.  The set of aligned null directions is
  described by a a system of alignment polynomials whose coefficients
  are derived from the components of the tensor.  The algebraic
  properties of the alignment polynomials can be used to classify the
  corresponding tensors and to put them into normal form.  The
  alignment classification paradigm is illustrated with a discussion
  of bivectors and of Weyl-type tensors.  Note: an earlier version of
  this manuscript was published in the proceedings of SPT 2004.  The
  present version has been expanded to include a discussion of
  complexified alignment.  Section 4 also corrects errors contained in
  the earlier manuscript.
\end{abstract}

\maketitle

\section{Introduction.}
We consider the algebraic aspects of the equivalence and
classification problems in Lorentzian geometry.  Let $U$ be a
representation of the Lorentz group $\cG$.  The classification problem
seeks a description of the orbit space $U/G$, and normal forms for the
orbit representatives.  The equivalence problem is the following
question: given of $u,\hu\in U$, does there exist a $g\in G$ such that
$\hu = g\cdot u$?

In 4 dimensions,  tensor classification is important for physical
applications, and in particular for the study of exact solutions of
the Einstein equations\cite{kramer}.  Beyond the classical theory,
there has been great interest in higher dimensional Lorentz manifolds
as models for generalized field theories that incorporate
gravity\cite{wesson}.  The problem of tensor classification in higher
dimensions\cite{cmpppz,desmet} is therefore of interest.

Rank 2 tensors are can be classified using eigenvalues and by
considering the algebraic properties of the associated characteristic
polynomial.  In this paper we introduce the more general notion of an
aligned null direction --- a kind of Lorentzian eigenvector concept,
but valid for arbitrary tensor ranks.  In analogy with the
characteristic polynomial, aligned null directions are the zeros of
certain corresponding alignment polynomials.  However, unlike the
invariant, univariate characteristic polynomial, the alignment
polynomials are multivariate covariants.  The classification and
equivalence analysis proceeds by considering the alignment polynomials'
invariant algebraic properties.

Previously, alignment polynomials in 2 variables were used by Penrose
and Rindler\cite[Ch.  8]{penrind} for the classification of maximally
symmetric spinors in four dimensions.  The Penrose-Rindler approach
can be generalized by noting that the components of a Lor\-entz\-ian
tensor, of any symmetry type and in any dimension, can be naturally
ordered according to boost weight.  In essence, the $\vn$ of the NP
tetrad is counted with with weight $1$, the $\vell$ with weight $-1$,
and the space-like components with weight $0$.  We will call a null
direction $\vell$ aligned with the tensor, if the components with the
largest weight vanish along that direction.  In 4D, the zero set of
the Penrose-Rindler polynomial is just the locus of aligned null
directions.  

We illustrate alignment-based classification by discussing bivectors
and Weyl-type, rank 4 tensors.  We provide a complete classification
of bivectors in all dimensions, and give normal forms.  For Weyl-type
tensors, we show that the 4D PND equation is equivalent to the
alignment equations and hence has meaning in higher dimensions.
However, we show that for $N>4$, generically, these equations have no
solutions.  For more on the classification of higher-dimensional Weyl
tensors see \cite{cmpp04}.  A discussion of some more theoretical
aspects of alignment can be found in\cite{mcpp04}.  We note that a
similar notion of an aligned null directions, valid for tensor
products of multivectors but defined in terms of tensorial
contractions, was previously introduced and discussed in
\cite{pozoparra}.

\section{Alignment.}
Let $V$ be an $N$-dimensional vector space, either real or complex,
equipped with a non-degenerate inner-product $\eta_{ab}$.  When
working over the reals, we will assume that $\eta_{ab}$ has Lorentz
signature. We define a \emph{null frame} to be a basis
$\vell=\vm\subp{0}$, $\vn=\vm\subp{1}$,
$\vm\subp{2},\ldots,\vm\subp{N-1}$, satisfying $\eta_{ab} \ell^a n^b =
\eta_{ab} m\subp{i}{}^a m\subp{i}{}^b = 1$, with all other products
vanishing.  Throughout, Roman indices $a,b,c, \rA, \rB, \rC$ range
from $0$ to $N-1$.  Lower case indices indicate an arbitrary basis,
while the upper-case ones indicate a null frame.  Space-like indices
$i,j,k$ also indicate a null-frame, but vary from $2$ to $N-1$ only.
Thus, $\eta_{01}=\eta_{ii}=1$ and $\eta_{\sA\sB}=0$ for all other
choices of $\rA, \rB$. We raise and lower indices using $\eta_{ab}$;
in particular for space-like indices we have $\vm\subp{i} =
\vm\supp{i}$.  The Einstein summation convention is observed
throughout.  A vector $u^a$ will be called null, time-like, or
space-like if $u^a u_a$ is, respectively, zero, negative, or positive.

We let $\cG\subset\mathop{GL}(V)$ denote the group of inner-product
preserving transformations; these correspond to null-frame orthogonal
matrices
$$\eta_{\sA\sA'}\Phi\ud\sA\sB  \Phi\ud{\sA'}{\sB'}= \eta_{\sB\sB'}.$$
Also corresponding to each such automorphism is a 
change of null-frame
$\hvm\subp\sB = \vm\subp\sA \Phi\ud\sA\sB$.   Over the complex field,
$\cG$ is just the complex orthogonal group.  Over the reals, $\cG$ is the
group of Lorentz transformations. As generators of $\cG$
we make take null rotations about $\vn$ and $\vell$
\eqref{eq:nullrot} \eqref{eq:nullrotl}, boosts
\eqref{eq:boost}, and spins \eqref{eq:spin}:
\begin{gather}
  \label{eq:nullrot}
  \hvell= \vell
    +z^j \vm_j-\half z^j z_j\, \vn,\quad
    \hvn=  \vn, \quad
    \hvm_i=  \vm_i - z_i \vn;\\   
    \label{eq:nullrotl}
  \hvn= \vn
    +z^j \vm_j-\half z^j z_j\, \vell,\quad
    \hvell=  \vell, \quad
    \hvm_i=  \vm_i - z_i \vell;\\
  \label{eq:boost}
  \hvell=  \lambda\, \vell, \quad
  \hvn= \lambda^{-1}\vn, \quad
  \hvm_i=  \vm_i,\quad\lambda \neq 0;\\
  \label{eq:spin}
  \hvell=  \vell, \quad
  \hvn= \vn, \quad
  \hvm_j=  \vm_i\, X\ud{i}{j},\; X\ud{i}{j} X\du{k}{j} =
  \delta\ud{i}{k}.
\end{gather}

Let $\tT\in V^{\otimes p}$ be a rank $p$ tensor.  For a given list of
frame indices $\rA_1,\ldots,\rA_p$, we call the corresponding
$T_{\sA_1\ldots \sA_p}$ a \emph{null-frame scalar}.  
A change of frame transforms the scalars
according to
\begin{equation}
  \label{eq:scalarxform}
   \hat{T}_{\sB_1\ldots \sB_p}=
   T_{\sA_1\ldots \sA_p} \,
   \Phi\ud{\sA_1}{\!\sB_1} \cdots \Phi\ud{\sA_p}{\!\sB_p},\quad
\Phi\ud\sA\sB\in \cG.
\end{equation}
In particular, a boost \eqref{eq:boost}  transforms the scalars according to:
\begin{equation}
  \label{eq:boostxform}
  \hat{T}_{\sA_1\ldots \sA_p}= \lambda^{\bwt_{\sA_1\ldots\sA_p}}\,
  T_{\sA_1\ldots \sA_p},\quad  \bwt_{\sA_1\ldots\sA_p}=\bwt_{\sA_1}+\ldots+\bwt_{\sA_p},  
\end{equation}
where   $\bwt_0=1$, $\bwt_i=0$, $\bwt_1=-1$.
We will call $\bwt_{\sA_1\ldots\sA_p}$ \emph{the boost weight} of the
scalar $T_{\sA_1\ldots\sA_p}$.  Equivalently, the boost weight of
$T_{\sA_1\ldots \sA_p}$ is the difference between the number of
subscripts equal to $0$ and the number of subscripts equal to $1$.  

Let $[\vk]=\lspan \vk,\; k^a k_a=0$ be a null direction
and let $\vell, \vn,\vm_i$, be an arbitrary null-frame such that
$\vell$ is a scalar multiple of $\vk$. We define $\bo(\vk)$, the
\emph{boost order along $\vk$}, to be the maximum of all
$\bwt_{\sA_1\ldots\sA_p}$ for which $T_{\sA_1\ldots\sA_p}\neq 0$.  A
null rotation about $\vell$ \eqref{eq:nullrotl} fixes the leading
terms of a tensor, while boosts \eqref{eq:boost}\eqref{eq:boostxform}
and spins \eqref{eq:spin} subject the leading terms to an invertible
transformation \eqref{eq:scalarxform}.  It follows that the boost
order does not depend on a choice of a particular null-frame, but
rather on the choice of null direction spanned by $\vk$.  Therefore,
the definition of $\bo(\vk)$ is sound; the boost order is the same for
all null frames for which $\vell\in \lspan\vk$.

Finally, we let $\maxbo$ denote the maximum value of $\bo(\vk)$ taken over all
null vectors $\vk$, and say that a null vector $\vk$ is \emph{aligned}
with the tensor $\tT$ whenever $\bo(\vk)< \maxbo$.  In other words, an
aligned null direction is one for which the leading boost-weight
scalars vanish.
The value of $\maxbo$ depends on the rank and on the symmetry
properties of the tensor $\tT$.  Generically, for a rank $p$ tensor,
$\maxbo=p$.  However, if the tensor has some index skew-symmetry, then
$\maxbo$ will be smaller than $p$.

Algebraically special tensors can be characterized in terms of the
existence of aligned vectors, with increasing specialization indicated
by a higher order of alignment.  In a nutshell, one tries to normalize
the form of the tensor by choosing $\vell$ and $\vn$ so as to induce
the vanishing of the largest possible number of leading and trailing
null-frame scalars. The tensor can then be categorized by the extent
to which such a normalization is possible.

\begin{definition}
  Let $\tT$ be a rank $p$ tensor, and let $\vell$ be an aligned vector
  whose order of alignment is as large as possible. We define the
  \textbf{primary alignment type} of the tensor to be
  $\maxbo-\bwt(\vell)$. If there are no aligned directions, i.e., the
  alignment equations are over-determined, we will say the primary
  alignment type is 0.\footnote{In some cases, but not always, it is
    appropriate to refer to such tensors as type G, for {\bf general
      type}.}
\end{definition}

Supposing that an aligned $\vell$ does exist, we let $\vn$ be a
null-vector of maximal alignment, but subject to the constraint
$n^al_a=1$. \
\begin{definition}
  We define the \textbf{secondary alignment type} of the tensor to be
  $\maxbo-\bwt(\vn)$, and define the \textbf{alignment type} of the
  tensor to be the pair consisting of the primary and the secondary
  alignment type.
\end{definition}
If $\vell$ is the unique aligned direction, i.e.  if no aligned $\vn$
exists, then we define the alignment type to be the singleton
consisting of the primary alignment type. We will also speak of
complex and real alignment type, according to whether our setting is a
real or a complex inner-product space.

\begin{example}
For a bivector $K_{ab} = -K_{ba},$ we have $\maxbo=1$;
the corresponding boost weights are shown below. An aligned null
direction corresponds to $K_{0i}=0$.
\begin{equation}
  \label{eq:bvbw}
  K_{ab} = \overbrace{2K_{0i}\, n_{[a} m^i{}_{b]}}^1 +
  \overbrace{2K_{01}\, n_{[a}\ell_{b]}+ K_{ij}\, m^i{}_{[a}
    m^j{}_{b]}}^0+
  \overbrace{2K_{1i}\, \ell_{[a} m^i{}_{b]}}^{-1}.  
\end{equation}
The following alignment types are possible for a non-zero bivector $K_{ab}$:
type $(1,1)$ if it is possible to set $K_{0i}=K_{1i}=0$; type $2$
if it is possible to set $K_{0i}=K_{01}=K_{ij}=0$; type $1$ if it is
possible to set $K_{0i}=0$, but not possible at the same time to set
$K_{1i}=0$; and type 0 if $K_{0i}\neq0$ relative to every null frame.
\end{example}

\begin{example}
Consider a symmetric, rank $2$ tensor
 $S_{ab}=S_{ba}$.  In this case, $\maxbo=2$
 with exactly one scalar of maximal boost weight, namely $S_{00}$.  A null direction spanned by a null $k^a$ is aligned if and
 only   
 $$S_{ab} k^{a} k^{b}=0.$$
 Thus, the aligned null directions are given
 by the zeros of the corresponding $N$-variate, homogeneous quadratic
 polynomial restricted to the null-cone. 
\end{example}

\section{Alignment polynomials.}
We now show that the set of aligned directions is a variety, the zero
set of a finite number of polynomial equations.
The set of all null directions  is an $N-2$ dimensional
variety:
\[
\spPK^{N-2}=\{[\vk]: k^a k_a=2 k_0 k_1 + k_i k^i=0\}.
\]
Affine coordinates $z_i = k_i/k_1$ are defined for every choice of
null-frame.  Over the real field, we regard $[\vn]$ as a point at
infinity, and identify $\spRPK^{N-2}$ with real extended space
$\hreals^{N-2}=\Rset^{N-2} \cup \{\infty\}$, the one point
compactification of $\Rset^{N-2}$ homeomorphic to the sphere
$\spS^{N-2}$.  Complexified extended space $\hcnums^{N-2}$ is the
union of $\Cset^{N-2}$ with points at infinity having, respectively,
the form $[z^i \vm_i+\vn]$, and $[z^i\vm_i]$, where $z^i z_i=0$.

Let $\tT$
be a rank $p$ tensor and $\vm_\sA$ a null-frame.  For every choice of
indices $\rA_1,\ldots,\rA_p$ we define the polynomial
\begin{gather}
  \label{eq:nrpoly}
  p_{\sA_1\ldots\sA_p}(z_i) = T_{\sB_1\ldots \sB_p}\,
  \Lambda\ud{\sB_1}{\!\!\sA_1}(z_i)\cdots
  \Lambda\ud{\sB_p}{\!\!\sA_p}(z_i),
  \quad\text{where}\\
  \label{eq:nrxform}
  \Lambda\ud\sA\sB(z_i)=
  \begin{pmatrix}
    1 & 0 & 0 \\
    -\half z^j z_j & 1 & -z_i \\
    z^j & 0 & \delta\ud{j}{i}
  \end{pmatrix}
\end{gather}
is the matrix corresponding to a null rotation about $\vn$ (c.f.
equation \eqref{eq:nullrot}) with the parameters $z_i$ considered as
indeterminates.

By definition, a null vector $\vk = \vell - \half \zeta^i \zeta_i\vn
+\zeta^i \vm_i$, is aligned with $\tT$ if and only if $z_i=\zeta_i$
is a solution of the corresponding  \emph{alignment equations}
\begin{equation}
  \label{eq:aligneqs}
  p_{\sA_1\ldots\sA_p}(z_i)=0,\quad
  \bwt_{\sA_1\ldots\sA_p}= \maxbo.
\end{equation}  
Henceforth, we will refer to the $p_{\sA_1\ldots\sA_p}(z_i),\;
\bwt_{\sA_1\ldots\sA_p}= \maxbo$ as the alignment polynomials
corresponding to the tensor $\tT$.

Of course, the alignment polynomials are only defined up to a choice
of a null frame, and undergo a certain covariant transformation when
the frame is changed.
A $\cG$-transformation $\hvm_\sB = \vm_\sA \Phi\ud\sA\sB$ induces a
change of affine coordinate, a birational
transformation 
\begin{gather}
  \label{eq:mobxform}
  z^j = \frac{\phi^j(\hz_i)}{\phi^0(\hz_i)}, \quad\text{where} \\
  \label{eq:phidef}
  \phi^\sA(\hz_i) = \Phi\ud{\sA}0   +
  \Phi\ud\sA{i}\,\hz^i-\half\Phi\ud\sA1\, \hz^i \hz_i.  
\end{gather}
The form for the transformation \eqref{eq:mobxform} follows from the
relation
\begin{equation}
  \label{eq:framerel}
  \hvell - \half \hz^i \hz_i \hvn + \hz^i \hvm_i =
  \phi^0(\hz_i) \vell   + \phi^1(\hz_i)\vn  + \phi^j(\hz_i) \vm_j\, .
\end{equation}
Let us also note that a real transformation of form
\eqref{eq:mobxform} is a conformal transformation of $\spS^m$, and is
known as a M\"obius transformation\cite{beardon}. In terms of the
affine coordinates, null rotations about $\vn$ correspond to
translations; null rotations about $\vell$ to origin-fixing
inversions; boosts correspond to dilations; spins correspond to
rotations.

\begin{proposition}
  \label{prop:align}
  Let $\hvm_\sB=\vm_\sA\Phi\ud\sA\sB$ be two  null frames
  related by a  transformation in $\cG$.  The corresponding polynomials
   \eqref{eq:nrpoly} are  related by
  \begin{gather}
    \label{eq:eqcovariance}
    \hp_{\sA_1\ldots\sA_p}(\hz_i) = p_{\sB_1\ldots\sB_p}(z_i)
    \Upsilon\ud{\sB_1}{\sA_1}(\hz_i)  \cdots
    \Upsilon\ud{\sB_p}{\sA_p}(\hz_i),\quad \text{where},\\
    \label{eq:upsilondef}
    \Upsilon\ud\sA\sB = 
    \begin{pmatrix}
      \phi^0 & \Phi\ud01  & \phi\ud0{,i} \\
      0 & \displaystyle \frac{1}{\phi^0} & 0 \\
      0 & \quad \displaystyle \Phi\ud{j}{1} - \Phi\ud01
      \frac{\phi^j}{\phi^0}\quad & \displaystyle \phi\ud{j}{,i} -
      \phi\ud{0}{,i} \frac{\phi^j}{\phi^0}
    \end{pmatrix},
    \end{gather}
   and where $\phi^\sA{}_{,i}$ denotes the
  partial derivative of $\phi^\sA$ with respect to $\hz_i$.
\end{proposition}
Note that $\Upsilon\ud\sA\sB=0$ for $\bwt_\sB<\bwt_\sA$, and hence
$\hp_{\sA_1\ldots\sA_p}$ depends only on $p_{\sB_1\ldots\sB_p}$ for
which $\bwt_{\sB_1\ldots\sB_p}\geq \bwt_{\sA_1\ldots\sA_p}$.  Hence,
two sets of alignment polynomials,
$$p_{\sB_1\ldots\sB_p}(z_i),\quad \hp_{\sA_1\ldots\sA_p}(\hz_i),\qquad
\bwt_{\sA_1\ldots\sA_p}=\maxbo,$$
defined
relative to different frames, are birationally related. Hence, on the
open set of finite points, $\phi_0(\hz_i)\neq 0$, the zeros of
$p_{\sB_1\ldots\sB_p}(z_i)$ and the zeros of the transformed
$\hp_{\sA_1\ldots\sA_p}(\hz_i)$ coincide.

\section{Bivectors.}
Let $K_{ab}\in\Lambda^2 V^*$ be a bivector (we use the inner product
to identify two-forms and bivectors).  Throughout, let $K\ud{a}{b}$
denote the corresponding skew-symmetric transformation of $V$. Let
$V_0\subset V$ denote the corresponding kernel, and $V_0^\perp$ the
corresponding orthogonal complement.  The skew-symmetry implies that
the latter is an invariant subspace.

As per the definition \eqref{eq:nrpoly}, the alignment polynomials for
a bivector are
$$p_{0j}(z_i) = K_{ab}\, \hell^a\, \hatm\du{j}{b},\quad j=2,\ldots,N-1,$$
where $\hvell, \hvm_j$ are defined in \eqref{eq:nullrot}.  Expanding
these expressions 
we obtain
\begin{equation}
  \label{eq:bv_apoly}
  p_{0j}(z_i) = K_{0j} +z^i K_{ij} -z_j K_{01} - \frac{1}{2} z^i z_i
  K_{1j}- z^i z_j K_{i1}.  
\end{equation}
Thus, the aligned null directions are the solution set of a system of
$N-2$ quadratic equations in $N-2$ variables.  


\begin{proposition}
  A null direction spanned by a $k^a$ is aligned
  with $K_{ab}$ if and only $k^a$ is a null eigenvector:
  $K\ud{a}{b} k^b = \lambda k^a,\quad k^a k_a=0.$
\end{proposition}
\begin{proof}
  Let a null vector $k^a$ be given; extend it to a complex null-frame
  $\vell=\vk,\, \vn, \vm_i$.  We note that $k^a$ is an eigenvector if
  and only if $K_{ab} k^b$ is orthogonal to $m\du{i}{a}$, i.e., if and
  only if $K_{0i}=0$ --- the condition of alignment.
\end{proof}

\begin{proposition}
  Let $\lambda\neq0$ be an eigenvalue of $K\ud{a}{b}$. Then,
  necessarily, the corresponding eigenvector is null. Furthermore,
  $-\lambda$ is also an eigenvalue, and the $\lambda,-\lambda$
  eigenspaces are null-complementary.
\end{proposition}

\begin{proposition}
  Over $\cnums$, the bivector alignment equations 
  \eqref{eq:bv_apoly}  admit a zero.
\end{proposition}
\begin{proof}
  If $V_0$ contains a null-vector, we are done.  If not, the
  restriction of $K\ud{a}{b}$ to $V_0^\perp$ is non-singular.  Since
  $\cnums$ is the ground field, this restriction possesses a non-zero
  eigenvalue. The corresponding eigenvector is null.
\end{proof}

It follows that over $\cnums$, we may without loss of generality
take $\vn$ to be an aligned direction.  Hence $K_{1j}=0,$ and the
alignment equations assume a linear form:
\begin{equation}
  \label{eq:bv_linalign}
  p_{0j}(z_i) = K_{0j} +z^i K_{ij} -z_j K_{01}=0.
\end{equation}
The above linear system \eqref{eq:bv_linalign} may or may not admit
solutions.  Consequently, over $\cnums$, all bivectors are either alignment
type (1,1) or alignment type 1.\footnote{In 4D relativity, these
  correspond to type I and type N bivectors \cite{Stewart}. To
  maintain consistency with this relativity nomenclature, these
  classes have also been referred to as type $I_i$ and type $I$,
  respectively\cite{hdvsi}} Respective canonical forms for these two
classes are given below:
\begin{align}
  \label{eq:bv11form}
  K_{ab} &= \lambda_0 n_{[a} \ell_{b]}+
  \!\!\sum_{p=1}^{\lfloor N/2\rfloor-1} \!\!
  \lambda_p\,
  m\ud{2p}{[a} m\ud{2p+1}{b]},\\
  \label{eq:bv1form}
  K_{ab} &= n_{[a} m\ud{N-1}{b]}+
  \!\!\sum_{p=1}^{\lfloor (N-3)/2\rfloor} \!\!
  \lambda_p\,
  m\ud{2p}{[a} m\ud{2p+1}{b]}.  
\end{align}
Examination of \eqref{eq:bv_linalign}  reveals that type (1,1) is
described by the 
equation 
$\det(K_{ij}-K_{01}\delta_{ij})\neq 0,$
and as such forms a dense
open subset of the vector space of all complex bivectors.
In other words, type (1,1),
complex bivectors are the generic case. 

Next, let us consider the classification over the reals.  Let $V_+,
V_-, \Vi$ denote the direct sum of, respectively, the positive,
negative, and imaginary eigenspaces of $K\ud{a}{b}$.  The subspace
$\Vi$ is even-dimensional, with a positive definite inner-product
restriction.  The subspaces, $V_+, V_-$ are either trivial, or
one-dimensional, null-complementary subspaces.
\begin{proposition}
\label{prop:evenbv}
If $N$ is even, the bivector equations \eqref{eq:bv_apoly} admit
a real zero.
\end{proposition}
\begin{proof}
  It suffices to consider the case where $V_+, V_-$ are trivial. We
  will show that $V_0$ contains a real null-vector.  Suppose not.  Then,
  $V_0^\perp= \Vi$, and hence $V_0$ is even-dimensional with signature
  $(N-\dim \Vi,1)$.  This is a contradiction.
\end{proof}

If $N$ is odd, then a skew-symmetric $K_{ab}$ is necessarily
singular.  Let 
$$\det(\lambda \delta\ud{a}{b} - K\ud{a}{b}) = \lambda^N - \sigma_1
\lambda^{N-1} + \ldots + \sigma_{N-1} \lambda,$$
be the characteristic
polynomial of the corresponding transformation.  The coefficients
$\sigma_1, \ldots, \sigma_{N-1}, \sigma_N$ are invariants of the
transformation $K\ud{a}{b}$, with the determinant, $\sigma_N=0$ as
noted above.  Equivalently, these are the symmetric elementary
polynomials applied to the eigenvalues of $K\ud{a}{b}$.
\begin{proposition}
  If $N$ is odd, the bivector equations \eqref{eq:bv_apoly} admit a
  real zero if and only if $\sigma_{N-1}\leq 0$.
\end{proposition}
\begin{proof}
  Suppose that $V_+, V_-$ are non-trivial, and hence that a
  null-eigenvector exists. The list of eigenvalues consists of
  paired complex conjugates,  a positive number, a negative number,
  and one or more zeros.  It follows that $\sigma_{N-1}\leq 0$.

  Thus, it suffices to consider the cases where $V_+, V_-$ are
  trivial. Now there are two possibilities: either $V_0$ contains a
  real null vector, or it doesn't.    In the first case, since the
  inner product on $\Vi$
  is positive-definite, $\dim V_0\geq 2$, and hence $\sigma_{N-1}=0$.
  The second case can only occur if $V_0$ is the span of a time-like
  vector.  In this case, no real, null eigenvector exists, and
  $\sigma_{N-1}>0$. 
\end{proof}

Thus, for odd dimensions $N$  there exist bivectors of
real alignment type 0 (the complex alignment type is (1,1)).  The
corresponding canonical form is
\begin{equation}
  K_{ab} = \lambda_0 \lp n_{[a} m^{N-1}_{\;b]}+\ell_{[a} m^{N-1}_{\;b]}\rp
  +  \!\!\sum_{p=1}^{(N-3)/2} \!\!
  \lambda_p\,
  m\ud{2p}{[a} m\ud{2p+1}{b]}.
\end{equation}

\section{Weyl-type tensors.}
We define a \emph{Weyl-like tensor} $C_{abcd}$ to be a traceless,
valence 4 tensor with the well-known index symmetries of the Riemann
curvature tensor:
\[ C_{abcd} = -C_{bacd} = C_{cdab},\quad
C_{abcd}+C_{acdb}+C_{adbc}=0,\quad C_{abc}{}^b=0.\] We let $\cW_N$
denote the vector space of $N$-dimensional Weyl-like tensors.  It
isn't hard to show that $\cW_N$ has dimension
$\frac{1}{12}(N+2)(N+1)N(N-3)$.
\begin{table}[h]
    \caption{Boost weight of  the Weyl scalars.}
    \label{tab:bw2}
    \begin{tabular}{|c|c|c|c|c|}
      \hline
      $2$ & $1$ & $0$ & $-1$ & $-2$\\
      \hline
      $C_{0i0j}$ &  $C_{010i}, C_{0ijk}$ & $C_{0101}, C_{01ij},
      C_{0i1j}, C_{ijkl}$ & $C_{011i}, C_{1ijk}$ & $C_{1i1j}$\\
      \hline
    \end{tabular}
\end{table}

The maximal boost weight for a Weyl tensor is given by $\maxbo=2$.
The Weyl alignment polynomials are given by
\begin{equation}
  \label{eq:weylalign}
  p_{0i0j}(z_i)=C_{abcd}\, \hell^a\, \hatm\du{i}{b}\, \hell^b\,
  \hatm\du{j}{d},  
\end{equation}
where $\hvell, \hvm_i$ are defined in \eqref{eq:nullrot}.  Since
$p_{0i0}{}^i=0$, the aligned directions are  the solution set
of a system of 
$\half N(N-3)=\half (N-2)(N-1)-1,$
fourth order equations in $N-2$
variables.

In 4D, the principal null directions of the Weyl-like
tensor are defined in terms of the so-called PND
equation\cite{penrind}:
\begin{equation}
  \label{eq:pndeq}
 k^b k_{[e} C_{a]bc[d}  k_{f]}k^c=0,\quad k^a k_a=0.
\end{equation}
It is easy to establish that the PND equations are just the
homogeneous form of the alignment equations\cite{mcpp04}.
\begin{proposition}
  For every dimension $N$, a null vector $k^a$ satisfies the PND equation
  \eqref{eq:pndeq} if and only if it is aligned with $C_{abcd}$.
\end{proposition}

For $N\geq 4$ we have
$\half N(N-3) \geq N-2$,
with equality if and only if $N=4$.  Thus, a four-dimensional
Weyl-like tensor always possesses at least one aligned direction (see
below).  For $N>4$, the number of equations is greater than the number
of variables, and hence, generically, the alignment equations are inconsistent.
\begin{theorem}
  \label{thm:nopnds}
  If $N\geq 5$, then the subset of complex, Weyl-type tensors without
  aligned null directions is a dense, open subset of $\cW_n$. 
\end{theorem}
In other words, the generic Weyl-like tensor in higher dimensions does
not possess any aligned null directions, not even complex aligned null
directions.

Let us now re-derive the well-known Petrov-Penrose classification of
real, 4-dimensional Weyl-like tensors using alignment.  To facilitate
the calculations we switch to the complexified, NP tetrad:
$\vell=\vm_0,\, \vn= \vm_1\, \vm_2,\, \vm_{2'} = \overline{\vm}_2,$
with $\ell^a n_a=1,\, m\du{2}{a} m_{2'a} =-1$.  A null rotation about
$\vn$ now takes the form
$$
\hvell= \vell +z'\, \vm_2+z\, \vm_{2'}+ z z'\, \vn,\quad \hvn= \vn,
\quad \hvm_2= \vm_2 +z\,\vn,\quad \hvm_{2'} = \vm_{2'}+z'\,\vn,$$
where $z'=\bar{z}$ for a real transformation.  Since $C_{abcd}$ is
trace-free, equation \eqref{eq:weylalign} gives that $p_{0202'}=0$.
We also note that $\partial \hvell/\partial z' = \hvm_{2}$, and
$\partial \hvm_{2}/\partial z'=0$.  It follows immediately that
$\partial p_{0202}/\partial z'=0.$ Hence,
\[
  p_{0202}(z,z')=p_{0202}(z)=
  C_{0202}(z-\zeta_1)(z-\zeta_2)(z-\zeta_3)(z-\zeta_4) 
\]
is a fourth degree polynomial of one complex variable.  Furthermore,
$p_{02'02'}$ is the complex conjugate of $p_{0202}$, and we deduce
that, generically, there are 16 complex aligned null directions:
$z=\zeta_p,\, z'=\overline{\zeta_q}$ , with $p,q=1,2,3,4$.
The real aligned null directions correspond to $z'=\bar{z}$.

We also see that a real $C_{abcd}$ is completely determined by the
polynomial $p_{0202}(z)$, and that a change of null-frame transforms
the latter by a M\"obius transformation.  The usual classification of
the 4-dimensional Weyl tensor now follows by considering the root
multiplicities of this polynomial.  We have derived all this directly
by means of the alignment paradigm; there was no need to invoke spinors.

\section{Acknowledgements}
The author was partially supported by an NSERC  grant.
Discussions with A. Coley, N. Pelavas, V. Pravda, A. Pravdov\'a, J.
Senovilla are gratefully acknowledged.


\begin{thebibliography}{10}
  
\bibitem{beardon} A. F. Beardon, {\em The Geometry of Discrete Groups}
  Springer-Verlag, (1983).
 
\bibitem{carmeli} M. Carmeli, {\em Group theory and general
    relativity}.  McGraw Hill, (1977).
 
\bibitem{cmpppz} A. Coley, R. Milson, N. Pelavas, V. Pravda, A.
  Pravdov{\'a}, and R. Zalaletdinov, {\em Phys. Rev. D} {\bf 67}
  104020, (2003).
  
\bibitem{cmpp04} A. Coley, R. Milson, V. Pravda, and A. Pravdov\'a,
  {\em Class. Quantum Grav.} {\bf 21} L35, (2004).
  
\bibitem{desmet} P. DeSmet, {\em Class. Quantum Grav.} {\bf 19} 4877--4895,
  (2002).
  
\bibitem{kramer} D. Kramer, H. Stephani, M. MacCallum, C. Hoenselaers,
  and E Herlt, {\em Exact solutions of Einstein's field equations}
  Cambridge University Press, (2003).

\bibitem{mcpp04}
R. Milson, A. Coley, V. Pravda, and A. Pravdov\'a,
 {\em gr-qc/0401010}.

\bibitem{hdvsi}
A. Coley, R. Milson, V. Pravda, and A. Pravdov\'a,
\emph{gr-qc/0410070}

\bibitem{wesson}
J. M. Overduin and P. S. Wesson,
 {\em Phys. Rep.} {\bf 283} 303--378, (1997).
 
\bibitem{penrind} R. Penrose and W. Rindler, {\em Spinors and
    Space-time, vol. II}.  Cambridge University Press, (1986).
  
\bibitem{pozoparra} J. M. Pozo and J. M. Parra, \textit{Class. Quantum
    Grav.}  \textbf{19} 967-983, (2002).
  
\bibitem{Stewart} J. Stewart, {\em Advanced general relativity},
  Cambridge University Press, (1990).

\end{thebibliography}
\end{document}